\documentclass[a4paper]{ESASPCS13Style}
\usepackage{epsfig}

\begin{document}

\title{Ultracool Subdwarfs: Metal-poor Stars and Brown Dwarfs Extending into
the Late-type M, L and T Dwarf Regimes}

\author{Adam J.\ Burgasser\inst{1}, J.\ Davy Kirkpatrick\inst{2}, \and
S\'ebastien L\'epine\inst{3}}
\institute{Dept.\ Physics \& Astronomy, UC Los Angeles, Los Angeles, CA 90095 USA; \and
Infrared Processing and Analysis Center, MS 100-22, California Institute of Technology,
Pasadena, CA 91101 USA; \and Dept.\ Astrophysics, Div.\ Physical Sciences,
American Museum of Natural History, Central Park West at 79th Street, New York, NY 10024 USA}

\maketitle

\begin{abstract}
Recent discoveries from red optical proper motion and wide-field
near-infrared surveys have uncovered a new population of ultracool
subdwarfs -- metal-poor stars and brown dwarfs extending into the late-type M, L and
possibly T spectral classes.  These objects are among the first low-mass
stars and brown dwarfs formed in the Galaxy, and are valuable tracers of
metallicity effects in low-temperature atmospheres.  Here we review the
spectral, photometric, and kinematic properties of recent discoveries.
We also examine L subdwarf classification,
and discuss how a large, complete sample of substellar halo subdwarfs
could probe the star formation history of the Galaxy's Population II.
We conclude by outlining a roadmap for future work to
more thoroughly explore this old and very low-mass population.

\keywords{Stars: low-mass, brown dwarfs --- Stars: Population II --- Subdwarfs }
\end{abstract}

\section{Cool Subdwarfs ca. Cool Stars 12}

Subdwarfs are metal-deficient stars, classically defined as lying below
the stellar main sequence in optical color-magnitude diagrams (\cite{kui39}).
These objects are in fact not subluminous but rather hotter (i.e., bluer in
optical colors) than equivalent mass
main sequence dwarfs, a consequence of their reduced metal opacity
(\cite{cha51,san59}).
Cool subdwarfs (spectral types sdK and sdM) are typically
found to have thick disk or halo kinematics,
and in the latter case are presumably relics of the early
Galaxy, with ages of 10--13 Gyr.  Because low mass subdwarfs have lifetimes far in excess of
the age of the Galaxy, they are important tracers of Galactic chemical history,
and are representatives of the first generations of star formation.

At the time of the Cool Stars 12 meeting in Boulder, Colorado, USA,
the coolest known subdwarfs (sd) and extreme subdwarfs (esd) --
[Fe/H] $\sim$ -1.2 and -2.0, respectively (\cite{giz97}) --
extended down to spectral types sdM7/esdM7, with effective temperatures
T$_{eff} \ga 3000$ K (\cite{sch99,leg00}).  As such, all
cool subdwarfs known at that time were hydrogen burning low-mass stars.
These objects had been predominantly
identified from optical, typically photographic plate, proper motion surveys,
including Luyten's Half-Second Catalog (LHS: \cite{luy79a})
and Two-Tenths Catalog (NLTT: \cite{luy79b}), and the APM Proper
Motion Survey (\cite{shz00}).  Photographic plate surveys have the advantage of broad temporal
baselines, as much as 50 years in the case of the APM survey,
allowing determinations
of proper motions down to 0$\farcs$008-0$\farcs$010 yr$^{-1}$.
The high space velocities
of halo subdwarfs cause them to stand out from the overwhelming
number of field stars with similar optical colors,
particularly through the use of reduced proper motion (RPM) diagrams (\cite{luy25,rei84,war72};
see Figure 6).

At the same time, well over one hundred even cooler, roughly solar metallicity,
dwarf stars and brown dwarfs had been identified in several wide field imaging surveys,
including members of the new spectral classes L and T
(\cite{kir99,me02a,geb02}; also see S.\ Leggett, these proceedings).
With T$_{eff}$s as low as 700 K (\cite{gol04}),
L and T dwarfs emit the majority of their luminous flux in the near-infrared;
hence, the success of recent wide-field near-infrared surveys such as
2MASS (\cite{cut03}), DENIS (\cite{epc97}), and SDSS (\cite{yor00}) in
uncovering them.  Late-type M, L and T dwarfs have extremely red optical/near-infrared
colors ($R-J$ $>$ 5; \cite{dah02}).  Hence, these so-called {\em ultracool} stars
and brown dwarfs (defined here
as objects with spectral classes M7 and later; see \cite{kir97}) are generally
missed in photographic plate surveys.

\section{To the End of the sdM Sequence}

New proper motion surveys based on photographic plates have
recently pushed our compendium of cool subdwarfs to the end of the
M spectral class.  The surveys include the SUPERBLINK catalog
(\cite{lep02,lep03d}) and the SuperCosmos Sky Survey
(SSS; \cite{ham01a}; 2001b; 2001c).
The first has identified faint, high proper motion stars at
northern declinations through image analysis of
Digital Sky Survey scans of red photographic
plates from the POSS-I (\cite{abe59})
and POSS-II (\cite{rei91}) sky surveys.
New discoveries (the ``LSR'' stars) include four subdwarfs as
cool or cooler than the prior late-type record-holder, the sdM7 LHS 377 (\cite{giz97}),
and a handful of extreme subdwarfs within a subclass of the
esdM7 APMPM 0559-29 (\cite{sch99}).
The SSS catalog also makes
use of digitized data from POSS-I and POSS-II,
but includes additional data from the ESO
and UK Schmidt photographic surveys, including infrared $I_N$ (0.7-0.9 $\mu$m)
plates from the UK Schmidt survey (\cite{har81}).  A recent catalog of 11,289
proper motion stars from the SSS survey around the southern Galactic cap
is given in \cite*{por04}.
Besides the discovery of the closest brown dwarfs to the Sun (\cite{sch03}),
two ultracool M subdwarfs have been identified amongst the ``SSSPM stars'',
including the latest-type M subdwarf known, the sdM9.5 SSSPM 1013$-$13 (\cite{sch04}; also see
H.\ Zinnecker, these proceedings).

\begin{figure}[ht]
  \begin{center}
    \epsfig{file=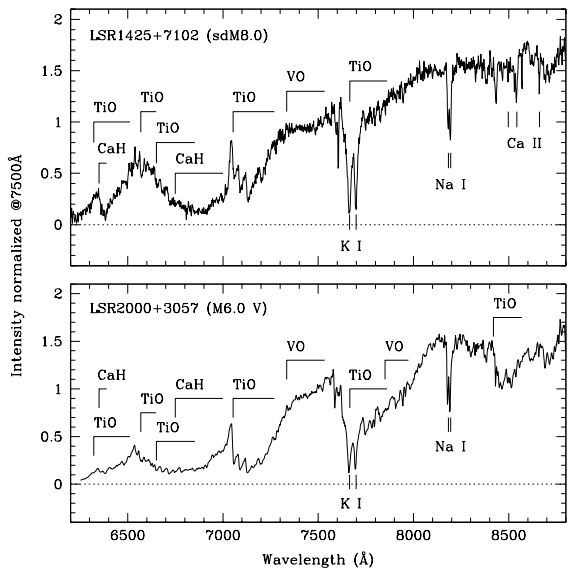, width=8cm}
  \end{center}
\caption{Optical spectra of the sdM8 LSR 1425+71 (top)
and the solar metallicity M6 V field star LSR 2000+30 (bottom).
These spectra demonstrate key metallicity
diagnostics at optical wavelengths, including enhanced
metal hydride bands (e.g., CaH bands at 6400 and
6800 {\AA}), weakened oxide bands (e.g., TiO and VO bands
at 6500, 7800 and 8500 {\AA}), and strong atomic lines.
(From L\'epine et al.\ 2003c).
\label{fig1}}
\end{figure}

Figure 1 (from \cite{lep03c}) shows the optical spectrum of one of the
new ultracool M subdwarfs, the sdM7.5 LSR 1425+71, compared
to a solar metallicity disk M dwarf.  Contrasting features between
these spectra demonstrate some of the metallicity diagnostics present
at red optical wavelengths.
Like normal disk dwarfs, absorption from molecular bands and neutral
atomic species dominate the optical spectra of M subdwarfs,
including the classic TiO and VO bands.  However,
M subdwarfs tend to have somewhat weaker metal oxide bands (note
the TiO and VO bands at 6500, 7800 and 8500 {\AA}) and stronger metal hydride
bands (note the strong CaH bands at 6400 and 6800 {\AA}) than disk dwarfs
of the same color, a fact known
since the 1960s (\cite{egg65}).  This shift in band strengths is likely
due to the competition between metal-hydride and metal-metal molecular
formation, which is skewed toward the former species in the metal-poor, hydrogen-rich
atmospheres of subdwarfs (\cite{mou76}).
The comparison between CaH and TiO bands forms the basis for the
M subdwarf classification scheme of \cite*{giz97}, the scheme by which
nearly all cool M subdwarfs known are classified (See $\S$ 5).
Other features that distinguish subdwarfs in the optical include
apparently enhanced atomic species, which emerge from the reduced metal oxide
opacity (note the Ca II lines around 8500 {\AA} in LSR 1425+71),
and bluer red optical slopes.

\begin{figure}[ht]
  \begin{center}
    \epsfig{file=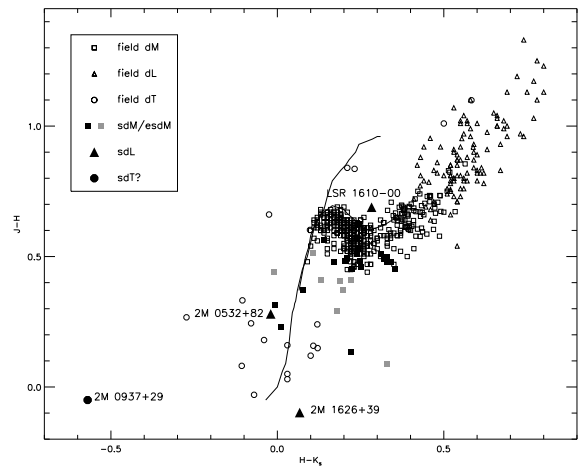, width=8cm}
  \end{center}
\caption{$JHK_s$ color-color diagram of M ({\em squares}), L ({\em triangles}),
and T ({\em circles}) dwarfs, comparing solar metallicity field objects ({\em open symbols})
to metal-poor subdwarfs and extreme subdwarfs ({\em filled symbols}).
M dwarf colors are from Leggett (1992) and Gizis et al.\ (2000);
colors for L and T dwarfs with small photometric errors
(${\sigma}_J < 0.05$ and 0.07 mag, respectively) are from 2MASS.
The Bessell \& Brett (1988) dwarf and giant tracks are superimposed for comparison.
While late-type M and L dwarfs becoming increasing red in near-infrared color
due to warm condensate dust radiation, subdwarfs
are increasingly blue due to
H$_2$ absorption.  Note in particular the L subdwarfs 2MASS 1626+39
and 2MASS 0532+82 and the possible T subdwarf 2MASS 0937+29.
\label{fig2}}
\end{figure}

At longer wavelengths, collision-induced H$_2$ absorption centered near
2.1 $\mu$m (1-0 quadrupole; \cite{sau94}) becomes an increasingly
important absorber in ultracool subdwarfs  due
to the reduction of competing opacities from metal-rich species.
This broad absorption feature gives cool subdwarfs
bluer near-infrared colors than equivalently-typed field
dwarfs, as illustrated in Figure 2.  CO and H$_2$O bands in the near-infrared are also observed
to be weaker in subdwarf spectra as compared to equivalently
classified field dwarfs (\cite{leg00,me04b}).

\section{L Subdwarfs}

The first unambiguous L subdwarf, 2MASS 0532+82, was serendipitously identified
in the 2MASS catalog.
This high proper-motion source
($\mu$ = 2$\farcs$6 yr$^{-1}$) has blue near-infrared colors
($J-K_s$ = 0.26) and no optical counterpart in POSS red
plates.  While its photometric properties initially made it a solid T dwarf candidate,
the observed spectrum of 2MASS 0532+82 (Figure 3) proved to be quite unlike
any T dwarf identified to date.
Its optical and $J$-band spectra closely resemble that of the L7
field dwarf DENIS 0205$-$11 (\cite{del97,kir99,mcl03}), with
strongly pressure-broadened fundamental doublets of Na I and K I (\cite{bur00}),
strong H$_2$O absorption in the optical and near-infrared,
metal hydride bands, and a steep red optical
slope.  However, metal hydride absorptions are significantly enhanced
in 2MASS 0532+82.
This source even exhibits TiH absorption at 9400 {\AA},
the first identification of the infrared band of this molecule
in an astrophysical source\footnote{Optical bands of TiH between 5200-5400 {\AA}
had previously been identified in the spectra of M giants (\cite{yer79}).}
(\cite{and03,me04b}; M.\ Cushing, priv.\ comm.).
Beyond 1.3 $\mu$m, the spectrum
of 2MASS 0532+82 deviates from that of DENIS 0205$-$11, with a blue,
relatively featureless slope and a notable absence of the 2.3 $\mu$m CO band.
The blue slope arises from H$_2$ absorption, which is notably enhanced in
this object.

\begin{figure*}[!ht]
  \begin{center}
    \epsfig{file=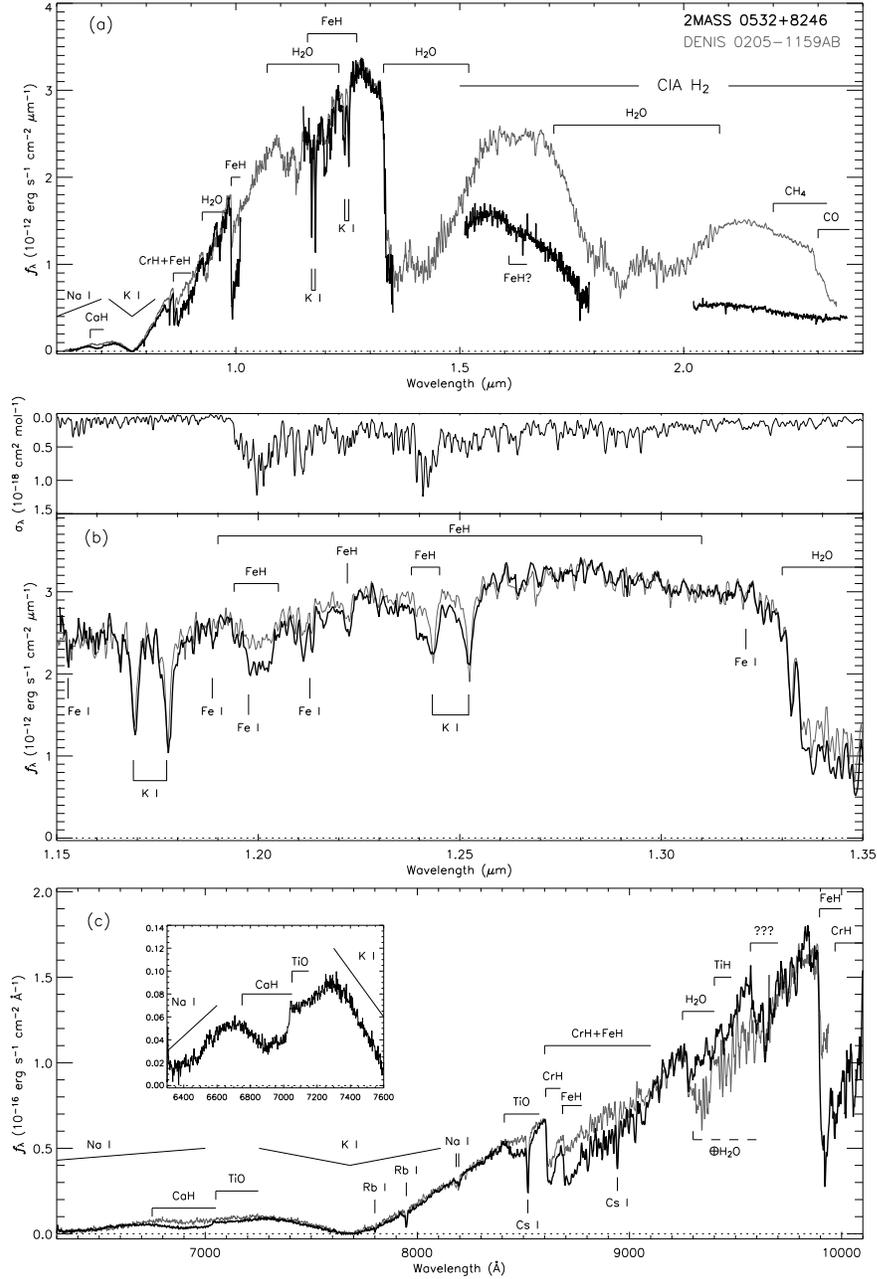, width=12cm}
  \end{center}
\caption{The spectrum of 2MASS 0532+82 (thick black line) as compared
to the L7 field dwarf DENIS 0205$-$11 (thin grey line).
In all panels, spectral data for 2MASS 0532+82 have been
shifted by its radial velocity ($V_{rad} = +195$ km s$^{-1}$),
and data for both sources are normalized at 1.27 $\mu$m.
(a) Observed 0.63--2.35 $\mu$m spectrum,
with NIRSPEC bands scaled to 2MASS photometry.
(b) {\em Top:} FeH opacity spectrum.
{\em Bottom:}
J-band spectrum of 2MASS 0532+82,
with line identifications for K I, Fe I, FeH, and H$_2$O.
(c) Red optical spectrum, with key features indicated.
Inset window shows a close-up of the 6350--7600 {\AA} spectral region,
highlighting strong CaH and weak TiO bands; no Li I or H$\alpha$ lines
are seen. (From Burgasser et al.\ 2003a)
\label{fig3}}
\end{figure*}

The observed spectral peculiarities of 2MASS 0532+82
are analogous to the metallicity features seen in
ultracool M subdwarf spectra, but extrapolated into the L dwarf regime.
This argues that 2MASS 0532+82 is simply a metal-poor L subdwarf.
The halo kinematics of 2MASS 0532+82 ($V_{tan}$ $\approx$
320 km s$^{-1}$; F.\ Vrba, private communication)
provide strong support for this interpretation.
One possible discrepancy, however,
is the presence of TiO absorption at 7050 and 8400 {\AA},
bands which are absent in DENIS 0205$-$11.
However, as TiO is depleted in field L dwarfs by the formation of more metal-rich
condensate minerals such as perovskite and enstatite
(\cite{lod02}), the retention of TiO gas may indicate that this process
is inhibited in metal-poor environments.
Chemical equilibrium calculations in subsolar metallicity environments
are needed to examine this possibility.

Effective temperature estimates for 2MASS 0532+82, conservatively
1400 $\la$ T$_{eff}$ $\la$ 2000 K, places it below the hydrogen burning
minimum mass (HBMM) for a metal poor star, $\sim$0.08 M$_{\sun}$ for
$Z$ = 0.1$Z_{\sun}$ (\cite{bur01}).  This makes 2MASS 0532+82 the first
example of a substellar halo subdwarf, and the first direct empirical
evidence\footnote{Indirect evidence
for the existence of Population II brown dwarfs has previously been seen
in halo and globular cluster mass
functions, which are flat to slightly rising near the HBMM (\cite{dah95,giz99,pio99,dig03}).}
that brown dwarfs formed early in the Galaxy's history.

At roughly the same time that 2MASS 0532+82 was announced,
a second L subdwarf was identified in the LSR catalog,
the high proper motion
star ($\mu$ = 1$\farcs$46 yr$^{-1}$) LSR 1610$-$00.  As reported in \cite*{lep03a},
this source exhibits features that are in common -- and in conflict --
with both late-type M and early-/mid-type L dwarf spectra.
The spectral energy distribution of this source as derived from photometry
is more similar to the former, with red near-infrared colors ($J-K_s$ = 0.89)
similar to a M7-M8 field dwarf.  This stands in contrast to the weak TiO
bands beyond 7500 {\AA} and absence of optical VO bands.  Metal hydride bands
are present, and the 6800 {\AA} CaH band is particularly strong, while
FeH bands are actually weaker than those of mid-type L dwarfs.  Most surprising
are the presence of Rb I lines, which are only observed in the spectra of
mid- and late-type field L dwarfs.  The spectral features in LSR 1610$-$00 are
similar to those of the sdM8 LSR 1425+71, with the exception of the Rb I lines
and a steeper red optical spectral slope.
Based on these diagnostics, \cite*{lep03a} propose that this object is another,
perhaps earlier-type, metal-poor L subdwarf.

Recently, a third L subdwarf has been identified
in the 2MASS database, 2MASS 1626+39 (\cite{me04b}).  This object
was found in the same search sample from which 2MASS 0532+82 was selected.
The spectrum of this object appears to progress in a natural sequence
beyond the latest-type M subdwarfs (Figure 4), giving strong support
to the interpretation that this object is simply a cooler, metal-poor star.
Indeed, the discovery of three L subdwarfs in the proximity
of the Sun (all three have $d < 50$ pc) indicates that
such objects are not isolated oddballs but in fact a {\em bona-fide} class of low-temperature,
metal-deficient, low-mass stars and brown dwarfs.

\section{T Subdwarfs?}

With the existence of at least one substellar L subdwarf, it is likely that
other low-metallicity brown dwarfs have cooled into the
T dwarf regime. Currently, one source remains a viable candidate, the peculiar T6
2MASS 0937+29 (\cite{me02a}).  This T$_{eff}$ $\approx$ 900 K
source has very blue near-infrared
colors ($J-K_s$ = $-$0.6, compared to typical T6 colors $J-K_s$ $\sim$ 0), consistent with
enhanced H$_2$ absorption as
seen in late-type M and L subdwarfs.  Its optical spectrum exhibits the steepest
red optical slope observed to date, caused by enhanced K I line broadening;
while the 9900 {\AA} FeH band is
stronger than those seen in other late-type T dwarfs (\cite{me03d}).  Comparison
of empirical data to spectral models by \cite*{bur02} also supports
a subsolar metallicity interpretation for this object, particularly
when the shape of its 1 $\mu$m spectral peak is considered.

However, these spectral peculiarities may be due to a different reason --
surface gravity effects.  Both the pressure-broadened
K I red wing at optical wavelengths and collision-induced H$_2$
absorption in the near-infrared are pressure-sensitive, and therefore
gravity-sensitive, features.  Indeed, at least three other late-type T dwarfs,
out of roughly 60 currently known, have similar
spectral peculiarities (Burgasser et al., in prep.; also see S.\ Leggett, these proceedings).
Such a high proportion of metal-poor subdwarfs in the T dwarf
regime is inconsistent with the relative
numbers of M- and L-type subdwarfs amongst field counterparts
(generally 0.2-0.3\%; \cite{dig03}), arguing in favor of
surface gravity being the dominant effect.  However, the optical spectrum
of 2MASS 0937+29 cannot be fit with any solar metallicity model currently available
(\cite{me03d}).  Clearly,
further observational and theoretical work is required to disentangle the
effects of gravity and metallicity in T dwarfs and ascertain the physical properties
of 2MASS 0937+29.
Because of the low atmospheric temperatures required to form
the characteristic CH$_4$ bands that define the T class, any T subdwarf identified
would necessarily be a substellar object.

\section{Classification of Ultracool Subdwarfs}

The most widely-adopted classification of M-type subdwarfs
is that defined by \cite*{giz97}, which relies primarily on the strength of
CaH (as a temperature index) and TiO (as a metallicity index) bands
in the 6200-7400 {\AA} region.  As we progress to later spectral subtypes,
however, these spectral features become less useful.
TiO absorption weakens as this gas is sequestered into condensate
species (at least for field dwarfs); while the spectral energy distribution shifts to longer
wavelengths, resulting in little measurable flux at visible wavelengths.
Recognizing this shortcoming, \cite*{lep03b} added an additional
classification index, Color-M,
to sample the spectral slope between 6500 and 8100 {\AA} for M subdwarfs,
similar to pseudocontinuum indices used by Kirkpatrick et al.\ (1991,1999)
to classify M and L disk dwarfs.  However, the L\'epine
subdwarf scheme continues to rely primarily on the \cite*{giz97} TiO and CaH
indices, and is therefore inappropriate
for L subdwarfs.  As new examples of ultracool subdwarfs are identified
and observed, a revised optical classification
relying on features at longer wavelengths will be needed.
Diagnostics might include the 7900 and 8400 {\AA} bands of VO and TiO,
the 8600 and 9900 {\AA} bands of CrH and FeH, the 9400 {\AA} band of TiH,
Cs I and Rb I alkali lines,
and color indices sampling longer wavelengths.

\begin{figure}[ht]
  \begin{center}
    \epsfig{file=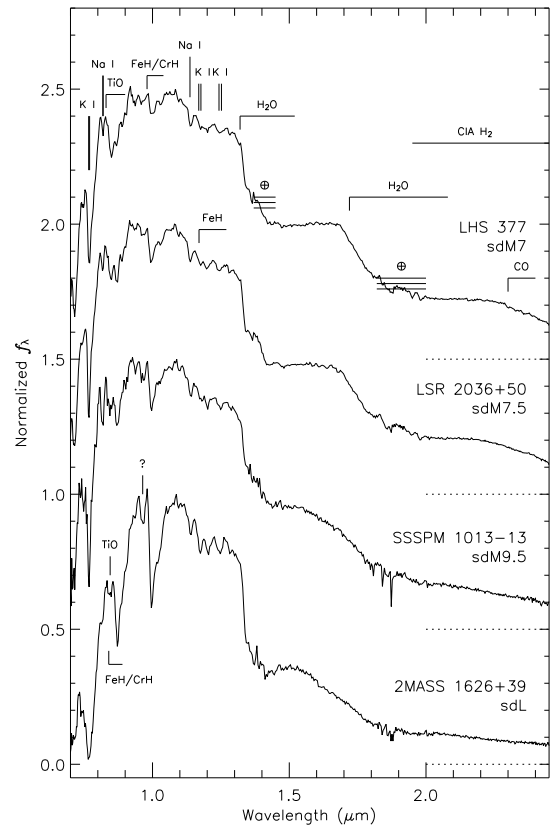, width=8cm}
  \end{center}
\caption{A low-resolution spectral sequence of late-type M subdwarfs
and the L subdwarf 2MASS 1626+39.  Note the strong metal hydride bands (e.g.,
FeH at 0.99 and 1.2 $\mu$m), weak metal oxides bands
(e.g., CO at 2.3 $\mu$m), and increasingly blue near-infrared
spectral slopes.  These spectra are quite unlike those of equivalently-typed
disk dwarfs, which tend to have strong metal oxides (e.g., TiO and VO),
weaker metal hydrides, and red near-infrared colors ($J-K > 1$).
(From Burgasser 2004b).
\label{fig4}}
\end{figure}

A logical step for future ultracool subdwarf classification is
to utilize spectral diagnostics at both optical and near-infrared
wavelengths.  Figure 4 illustrates a roadmap for this direction,
showing a low-resolution 0.7-2.5 $\mu$m
spectral sequence of ultracool subdwarfs from sdM7 to sdL (\cite{me04b}).
Many of the metallicity spectral diagnostics discussed above appear to vary
in a systematic way amongst the ultracool subdwarfs, including a weakening
of the 0.85 $\mu$m TiO and 2.3 $\mu$m CO bands; strengthening of FeH and CrH bands
at 0.86, 0.99, and 1.2 $\mu$m;
strengthening lines of K I
at 0.77, 1.17, and 1.25 $\mu$m; strengthening H$_2$O absorption at 1.3 $\mu$m;
and spectral slopes that become increasing red shortward of 1 $\mu$m
and increasingly blue at longer wavelengths.
The regular evolution of these features in the spectra of Figure 4
is promising for the future classification of ultracool subdwarfs.

Despite these optimistic signs,
it is important to stress that before a formal scheme can be codified,
more examples of these objects must be identified.  Accurate spectral classification
relies on the comparison of spectra to defined standard stars, and
not the simple extrapolation of spectral indices.

\section{Substellar Subdwarfs and Star Formation Rate}

Most ultracool subdwarfs identified to date
have kinematics consistent with membership in
the halo (Population II) or thick disk
(Intermediate Population II) of the Galaxy.  Hence, an accurate
assessment of the Luminosity Function (LF) of late-type M and L subdwarfs
enables us to examine the low-mass end of the Mass Functions (MF)
for these populations down to and below the HBMM.
Furthermore, as long-lived sources, ultracool subdwarfs
are excellent tracers of physical conditions in the early Galaxy, including chemical
enrichment.

\begin{figure}[ht]
  \begin{center}
    \epsfig{file=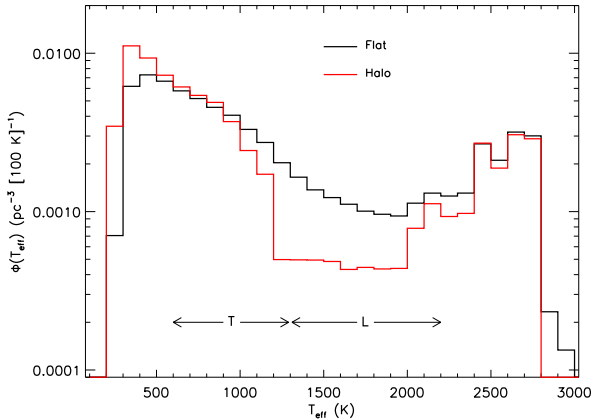, width=8cm}
  \end{center}
\caption{A comparison of simulated LFs of low-mass stellar and substellar
objects assuming a power-law MF (dN/dM $\propto$ M$^{-0.5}$)
and two underlying star formation rates: constant
over 10 Gyr (``flat'' distribution, top line)
and a 1 Gyr burst 10 Gyr in the past (``halo'' distribution, bottom line).
Solar metallicity is assumed for both populations, and the MFs are normalized
to have the same space density at 0.1 M$_{\sun}$.
The star formation history has a significant effect on the relative
numbers of 1200--2000 K objects, equivalent to L-type dwarfs, with
a third fewer sources in the halo distribution.
This deficiency is caused by the thermal evolution of brown dwarfs,
which pass through the L-dwarf phase in a few
Gyr.  The effect should be enhanced in a metal-poor halo population, which cool more
rapidly due to reduced atmospheric opacities.  Numbers of cooler (T-type) dwarfs
are unaffected by the underlying star formation history, hence comparison between these
two classes could provide
independent information on both the MF and star formation history.
(Adapted from Burgasser 2004a).
\label{fig5}}
\end{figure}

Measuring the LF of ultracool subdwarfs is currently an
intractable task due to
paucity of sources known.  However, a significant and complete sample of
substellar subdwarfs could provide information on both the
star formation history and MF of the Galactic halo and thick disk.
This is because substellar objects continually evolve over time to lower
luminosities and cooler effective temperatures due to the lack
of sustained hydrogen fusion.  As such, they make excellent chronometers.
Figure 5 diagrams an application of this fact, comparing simulated LFs (\cite{me04a})
for two populations: one with a constant star formation rate over the age of the Galaxy
(appropriate for a disk population) and one with a single 1 Gyr burst 10 Gyr in the
past (appropriate for a halo population).  In the T$_{eff}$ regime appropriate
for L-type dwarfs, there is a clear deficiency of objects in the halo population,
the result of the long-term thermal evolution of most brown dwarfs to cooler
temperatures. There is no such depletion amongst the T-type
dwarfs, and it has been shown that the LF of T dwarfs is highly
sensitive to the underlying MF while the LF of L dwarfs is not
(\cite{me04a,all04}).
Hence, by measuring the space densities of both populations, one can conceivably
measure both the MF and star formation history of a particular population.
While such a measurement may be several years off, Figure 5 demonstrates
how ultracool halo subdwarfs can ultimately be used to measure global Galactic properties.

\section{Ultracool Subdwarfs ca. Cool Stars 13}

Table 1 lists all ultracool subdwarfs known at (or soon after) the
Hamburg Cool Stars 13 meeting.  The shortness of this list betrays our limited
understanding of these metal-poor cool stars and brown dwarfs, and it is clear
that continuing work is needed to discover and characterize them.

\begin{table*}[!th]
  \caption{Known Ultracool Subdwarfs sdM7/esdM7 and Later.}
  \label{tab:table}
  \begin{center}
    \leavevmode
    \footnotesize
    \begin{tabular}[h]{llccccl}
      \hline \\[-5pt]
      Name & SpT$^a$  & $R-J$$^b$ & $J$$^b$ & $J-K_s$$^b$   &  $\mu$ (yr$^{-1}$) & Reference  \\[+5pt]
      \hline \\[-5pt]
      APMPM 0559$-$29 & esdM7  & 3.2  & 14.89  & 0.43  & 0$\farcs$38 & \cite*{sch99} \\
      LHS 377 & sdM7  & 4.2  & 13.19  & 0.71  & 1$\farcs$25 & \cite*{giz97} \\
      SSSPM 1930$-$43 & sdM7  & 3.6  & 14.79  & 0.70  & 0$\farcs$87 & \cite*{sch04} \\
      LSR 2036+50 & sdM7.5  & 3.9  & 13.61  & 0.68  & 1$\farcs$05 & \cite*{lep03b} \\
      LSR 1425+71 & sdM8  &  3.8 & 14.78  & 0.45  & 0$\farcs$64 & \cite*{lep03c} \\
      SSSPM 1013$-$13 & sdM9.5  & 4.4  & 14.59  & 0.24  & 1$\farcs$03 & \cite*{sch04} \\
      LSR 1610$-$00 & sdL$^c$  &  4.6 & 12.91  & 0.89  & 1$\farcs$46 & \cite*{lep03a} \\
      2MASS 1626+39 & sdL  & 5.4  &  14.44 & $-$0.03  & 1$\farcs$27 & \cite*{me04b} \\
      2MASS 0532+82 & sdL  & $>$ 5.5  &  15.18 & 0.26  & 2$\farcs$60 & \cite*{me03c} \\
      2MASS 0937+29 & sdT?$^d$  & $>$ 6  &  14.65 & $-$0.62  & 1$\farcs$62 & \cite*{me02a} \\
      \hline \\
      \end{tabular}
  \end{center}
{\scriptsize
$^a$M subdwarf spectral types based on the optical classification scheme of \cite*{giz97}; an L subdwarf
classification scheme has not yet been defined (see $\S$ 4). \\
$^b$$R$ magnitudes from the USNO B1.0 (\cite{mon03}); $JK_s$ magnitudes from 2MASS (\cite{cut03}).\\
$^c$Cushing et al.\ (in prep.) have questioned the subdwarf status of
LSR 1610$-$0040 as its near-infrared
colors and spectrum are atypical of other late-type sdMs and
sdLs.  Its spectrum is clearly inconsistent with that of a disk dwarf, however, and further
observational work is needed. \\
$^d$Subsolar metallicity effects observed in the spectrum of this object may be confused with
high surface gravity.  See discussion in $\S$ 3.
}
\end{table*}

One potentially powerful technique to identify new ultracool subdwarfs
that has yet to be adequately explored is a
near-infrared proper motion survey.  It is clear from
studies of late-type M, L and T dwarfs, and the few ultracool subdwarfs now known, that
the optical bands used in contemporary proper motion surveys
miss the peak of the spectral energy distributions of later-type objects.
Examination of Figure 4 shows that surveys conducted around 1 $\mu$m (e.g.,
$Z$, $Y$, and $J$ photometric bands) are optimally suited to detect ultracool subdwarfs,
as verified by the identification of two L subdwarfs in the 2MASS survey.
Red and infrared RPM diagrams, such as that illustrated in
Figure 6, could help distinguish those objects.
Work is currently in progress to match multiple-epoch observations in the 2MASS survey to
identify very high proper motion sources that are potential ultracool subdwarfs;
cross-correlation between the 2MASS, DENIS, and SDSS surveys could similarly identify
high proper motion, near-infrared bright sources.
However, these surveys sample a relatively
narrow timeframe, roughly 2-5 years, restricting work to the highest motion,
and hence most nearby, sources. A second epoch wide field survey
is needed to detect motion from more distant high velocity stars, enabling a larger
and more statistically complete sample.  With the development of large-format
near-infrared detectors, such surveys are possible and, indeed, already in
development.  These include the UKIRT Infrared Deep Sky
Survey\footnote{See http://www.ukidss.org/index.html.} (UKIDSS; S.\ Leggett, these
proceedings) and
the VISTA telescope\footnote{See http://www.vista.ac.uk/index.html.} (\cite{mcp03}).
By using existing sky surveys for first epoch astrometry, such
programs may ultimately produce ultracool subdwarf samples large enough to begin
more global studies, such as measurement of the low-mass halo MF.

\begin{figure}[!ht]
  \begin{center}
    \epsfig{file=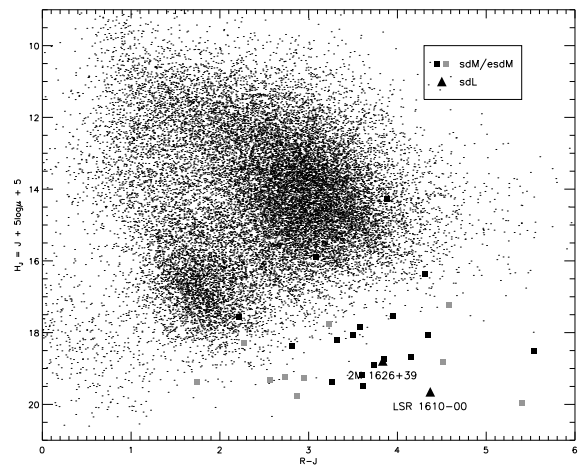, width=8cm}
  \end{center}
\caption{$J$-band RPM diagram for sources in the revised
NLTT catalog (Salim \& Gould 2003) and ultracool M ({\em squares}) and L ({\em triangles})
subdwarfs and extreme subdwarfs.  The RPM, defined as $H_J = J + 5\log{\mu} + 5$ =
$M_J + 5\log{V_{tan}} - 3.38$,
distinguishes field and halo stars by the higher tangential motions of
the latter.  This segregation is apparent amongst the bluer stars in the plot
(note the two populations separated by a dashed line; the field dwarfs are to the
right of the line), but can be somewhat muddled amongst the latest-type dwarfs.
Nonetheless, ultracool extreme subdwarfs tend to lie at fainter RPMs than subdwarfs
and field stars, while the latest-type objects tend to the lower right of the diagram,
indicative of lower luminosities, redder colors, and (in the case of subdwarfs) high
space velocities.
\label{fig6}}
\end{figure}

Additional observational work is required for the existing set of ultracool subdwarfs.
A subset of investigations currently possible
include optical and near-infrared spectroscopy to investigate spectral diagnostics
of temperature and metallicity; high-resolution spectroscopy to measure specific
abundances and radial velocities; broad-band imaging to measure bolometric corrections;
and parallax measurements to determine absolute
brightnesses, effective temperatures, and space motions.
Observations at longer wavelengths (e.g., using the Spitzer telescope) may
also provide new information on the atmospheric composition of these
objects, particularly in regards to the presence and constituency of dust
condensates.  High-resolution imaging and/or high-resolution spectroscopic
monitoring can probe multiplicity and binary distributions,
a clue to formation mechanisms.

These observational advances must also be matched by improvements in theoretical
evolutionary and spectral models for cool, metal-poor stars and brown dwarfs.
The most recent theoretical work includes
that of \cite*{bar97}, who have computed evolutionary models for subsolar metallicity stars
down to the HBMM; \cite*{leg00}, who have
compared observational spectra of M subdwarfs down to sdM7 (LHS 377) to theoretical
models developed by the Lyon group (\cite{hau99});
and \cite*{bur02}, who examined metallicity effects in theoretical spectra
of T dwarfs such as 2MASS 0937+29.
Clearly, both evolutionary and spectral models will need to be expanded to address the properties of
the recently discovered ultracool subdwarfs.  In addition,
chemical equilibrium models may need to be examined, as non-solar metallicities
should have a significant effect on the abundances of the most
spectrally active species in an ultracool subdwarf atmosphere, including
the condensate species that dominate L field dwarf spectra. In this context,
the interaction between observation and theory can be expected to be quite
fruitful for this new class of subdwarfs in the years to come.

\begin{acknowledgements}

The authors thank J.\ Gizis, J.\ Liebert, and I.\ N.\ Reid for their helpful
comments on the original manuscript.  A.\ J.\ B.\ acknowledges support provided by NASA through
Hubble Fellowship grant HST-HF-01137.01 awarded by the Space Telescope Science Institute,
which is operated by the Association of Universities for Research in Astronomy,
Incorporated, under NASA contract NAS5-26555.
S.\ L.\ is a Kalbfleich research fellow of the American Museum of Natural History.
This publication makes use of data from the Two
Micron All Sky Survey, which is a joint project of the University
of Massachusetts and the Infrared Processing and Analysis Center,
funded by the National Aeronautics and Space Administration and
the National Science Foundation.
2MASS data were obtain through
the NASA/IPAC Infrared Science Archive, which is operated by the
Jet Propulsion Laboratory, California Institute of Technology,
under contract with the National Aeronautics and Space
Administration.

\end{acknowledgements}

\end{document}